\documentclass[aps,showpacs,amsmath,floatfix]{revtex4}
\usepackage{graphicx}%
\usepackage{dcolumn}
\usepackage{color}
\usepackage{amsmath}
\usepackage{here}
\begin{document}
\newcommand{\be}{\begin{equation}}
\newcommand{\ee}{\end{equation}}
\newcommand{\rojo}[1]{\textcolor{red}{#1}}

\title{Stable Vortex Solitons in Nonlocal Self-Focusing Nonlinear Media}

\author{Alexander I. Yakimenko$^{1,2}$, Yuri A. Zaliznyak$^1$, and Yuri Kivshar$^{3}$}

\affiliation{$^1$Institute for Nuclear Research, Kiev 03680, Ukraine \\
$^2$Department of Physics, Taras Shevchenko National University,
Kiev 03680, Ukraine \\
$^3$Nonlinear Physics Center, Research School of Physical Sciences
and Engineering, Australian National University, Canberra ACT
0200, Australia}

\begin{abstract}
We reveal that spatially localized vortex solitons become {\em
stable} in self-focusing nonlinear media when the vortex
symmetry-breaking azimuthal instability is eliminated by a
nonlocal nonlinear response. We study the main properties of
different types of vortex beams and discuss the physical mechanism
of the vortex stabilization in spatially nonlocal nonlinear media.
\end{abstract}


\maketitle

Vortices are fundamental objects which appear in many branches of
physics~\cite{book}. In optics, vortices are usually associated
with phase singularities of diffracting optical beams, and they
can be generated experimentally in different types of linear and
nonlinear media~\cite{review}. However, optical vortices become
{\em highly unstable} in self-focusing nonlinear media due to the
symmetry-breaking azimuthal instability, and they decay into
several fundamental solitons~\cite{KivsharAgrawal}. In spite of
many theoretical ideas to stabilize optical vortices in specific
nonlinear media~\cite{theory}, no stable optical vortices created
by coherent light were readily observed in
experiment~\cite{remark}. Thus, the important challenge remains to
reveal physical mechanisms which would allow the first
experimental observation of stable vortices in realistic nonlinear
media.

In this Letter, we reveal that the symmetry-breaking azimuthal
instability of the vortex beams can be eliminated in a medium
characterized by a nonlocal nonlinear response. This observation
allows us to suggest a simple and realistic way to generate
experimentally the first stable spatially localized vortices in
self-focusing nonlinear media. We study the main properties and
stability of different types of vortex beams, and discuss the
physical mechanism of their stabilization in spatially nonlocal
nonlinear media.

We notice that there are many physical systems characterized by
nonlocal nonlinear response. In particular, a nonlocal response is
induced by heating and ionization, and it is known to be important
in plasmas~\cite{LitvakSJPlPhys75}. Nonlocal response is a key
feature of the orientational nonlinearities due to long-range
molecular interactions in nematic liquid
crystals~\cite{ContiPRL03}. An interatomic interaction potential
in Bose-Einstein condensates with dipole-dipole interactions is
also substantially nonlocal~\cite{BEC}. In all such systems,
nonlocal nonlinearity can be responsible for many novel features
such as the familiar effect of the collapse arrest
\cite{TuritsynTMF85,KrolikovskiJOptB04}.

We consider propagation of the electric field envelope $E(X,Y,Z)$
described by the paraxial wave equation:
\begin{equation}
\label{eq:FieldDimensional} 2ik_0 \frac{\partial E}{\partial Z}
+\frac{\partial^2 E}{\partial X^2}+\frac{\partial^2 E}{\partial
Y^2} +k_0^2n_T\Theta E=0,
\end{equation}
where $k_0$ is the wave number, and the function $\Theta$
characterizes a nonlinear, generally nonlocal, medium response.
For example, in the case of the wave beam propagation in partially
ionized plasmas, $\Theta=T_e'/T$ is the relative electron
temperature perturbation, $T$ being unperturbed temperature, the
coupling coefficient $n_T=+1$. Stationary temperature perturbation
obeys the equation~\cite{LitvakSJPlPhys75}:
\begin{equation}
\label{eq:TemperatureDimensional}
\alpha^2\,\Theta-l^2_e\left[\frac{\partial^2 \Theta}{\partial
X^2}+\frac{\partial^2  \Theta}{\partial Y^2}+ \frac{\partial^2
\Theta}{\partial Z^2}\right]=|E|^2/E_c^2,
\end{equation}
where $E_c^2=3mT (\omega_0^2+\nu_{e}^2)/e^2$, $\nu_e$ is the
electron collision frequency, $\omega_0$ is the wave frequency,
$m$ is electron mass, $\alpha^2$ characterizes the relative
portion of the energy that electron deliver to a heavy particle
during single collision. The second term describes {\em  thermal
diffusion} with the characteristic spatial scale $l_e^2$. Note,
that the model identical to Eqs.~(\ref{eq:FieldDimensional}) and
(\ref{eq:TemperatureDimensional}) has been employed in
Ref.~\cite{ContiPRL03} to study two-dimensional bright solitons
observed experimentally in nematic liquid
crystals~\cite{ContiPRL03,ContiPRL04}. In this case, the field
$\Theta$ describes the spatial distribution of the molecular
director.

Rescaling the variables, $(X, Y)=l_e (x, y)$, $Z=2l_ez/\epsilon$
and the fields, $E=(E_c\epsilon/\sqrt{n_T})\Psi(x,y,z)$ and
$\Theta=(\epsilon^2/n_T)\theta(x,y,z)$, where
$\epsilon=(k_0l_e)^{-1}$, we present
Eq.~(\ref{eq:TemperatureDimensional}) in the dimensionless form,
\begin{equation}
\label{eq:Temperature}
\alpha^2\,\theta-\Delta_\perp\theta-\frac{\epsilon^2}{4}\frac{\partial
^2\theta} {\partial z^2}=|\Psi|^2,
\end{equation}
where $\Delta_\perp=\partial^2/\partial x^2+\partial^2/\partial
y^2$ is the transverse Laplacian.  For the analysis performed
below, we omit in Eq.~(\ref{eq:Temperature}) the term proportional
to $\epsilon^2$.

Thus, the basic dimensionless equations describing the propagation
of the electric field envelope $\Psi(x,y,z)$ coupled to the
temperature perturbation $\theta(x,y,z)$ become
\begin{equation}
   \begin{array}{l} {\displaystyle
       i\frac{\partial \Psi}{\partial z}+\Delta_\perp\Psi +\theta\Psi=0,
       } \\*[9pt] {\displaystyle
\alpha^2\,\theta-\Delta_\perp\theta=|\Psi|^2.
   }\end{array}
   \label{eq:NLS}
\end{equation}
In the limit $\alpha^2\gg 1$, we can neglect the second term in
the equation for the field $\theta$ in Eq.~(\ref{eq:NLS}) and
reduce this system to the standard local nonlinear Schr\"{o}dinger
(NLS) equation with cubic nonlinearity. The opposite case, i.e.
$\alpha^2 \ll 1$, can be referred to as a strongly nonlocal regime
of the beam propagation.

We look for stationary solutions of the system (\ref{eq:NLS}) in
the form, $\Psi(x,y,z)=\psi(r) \exp(im\varphi+i\Lambda z)$, where
$\varphi$ and $r=\sqrt{x^2+y^2}$ are the azimuthal angle and
radial coordinate, respectively, and $\Lambda$ is the beam
propagation constant. Such solutions describe either the
fundamental optical soliton, when $m=0$,  or the vortex soliton
with the topological charge $m$, when $m \neq 0$.

The beam radial profile $\psi(r)$ and accompanied temperature
field $\theta(r)$ can be found by solving the system of ordinary
differential equations,
\begin{equation}
   \begin{array}{l} {\displaystyle
      -\lambda\psi+\Delta_r^{(m)}\psi+\theta\,\psi=0,
       } \\*[9pt] {\displaystyle
\alpha^2\theta-\Delta_r^{(0)}\theta=|\psi|^2,
   }\end{array}
   \label{eq:psi_r}
\end{equation}
where $\Delta_r^{(m)}=d^2/d r^2+(1/r)(d/d r)-(m^2/r^2)$, and we
rescale $\psi$, $\theta$, $1/r^2$, and $\alpha^2$ by the factor
$\Lambda$, and choose the unity propagation constant. Boundary
conditions are: for the localized vortex field, $\psi(\infty)
=\psi(0)=0$, and for the temperature field, $d\theta/dr|_{r=0}=0$
and $\theta(\infty) =0$.

The second equation of the system (\ref{eq:NLS}) can be readily
solved for axially symmetric intensity distribution $|\Psi|^2$,
\begin{equation}
\theta(r,z)=\int_0^{+\infty}|\Psi(\xi, z)|^2 G_0(r, \xi;
\alpha)\xi d\xi,
\label{eq:Theta(R)}
\end{equation}
where $G_0$ is the Green's function defined at $\nu=0$ from the
general expression
\begin{equation}
\label{eq:Green'sfunction}
G_\nu(\xi_1,\xi_2;
a)=\left\{
  \begin{array}{lc}
    K_\nu(a\xi_2)I_\nu(a\xi_1), & 0\le \xi_1<\xi_2, \\
    I_\nu(a\xi_2)K_\nu(a\xi_1), & \xi_2<\xi_1<+\infty,
  \end{array}
\right.
\end{equation}
and $I_\nu$ and $K_\nu$ are the modified Bessel functions of the
first and second kind, respectively. Thus, Eqs.~(\ref{eq:psi_r})
are equivalent to a single integro-differential equation obtained
from (\ref{eq:psi_r}) when $\theta(r)$ is eliminated, or to a
single integral equation:
\begin{equation}
\label{eq:E_rintegralEq}
\psi(r)=\int_0^{+\infty}\theta(\eta)
\psi(\eta)\,G_m(r,\eta;\sqrt{\lambda}) \eta\,d\eta,
\end{equation}
where $G_m$ is defined by Eq.~(\ref{eq:Green'sfunction}) and
$\theta$ is given by Eq.~(\ref{eq:Theta(R)}).

\begin{figure}
\includegraphics[width=3.4in]{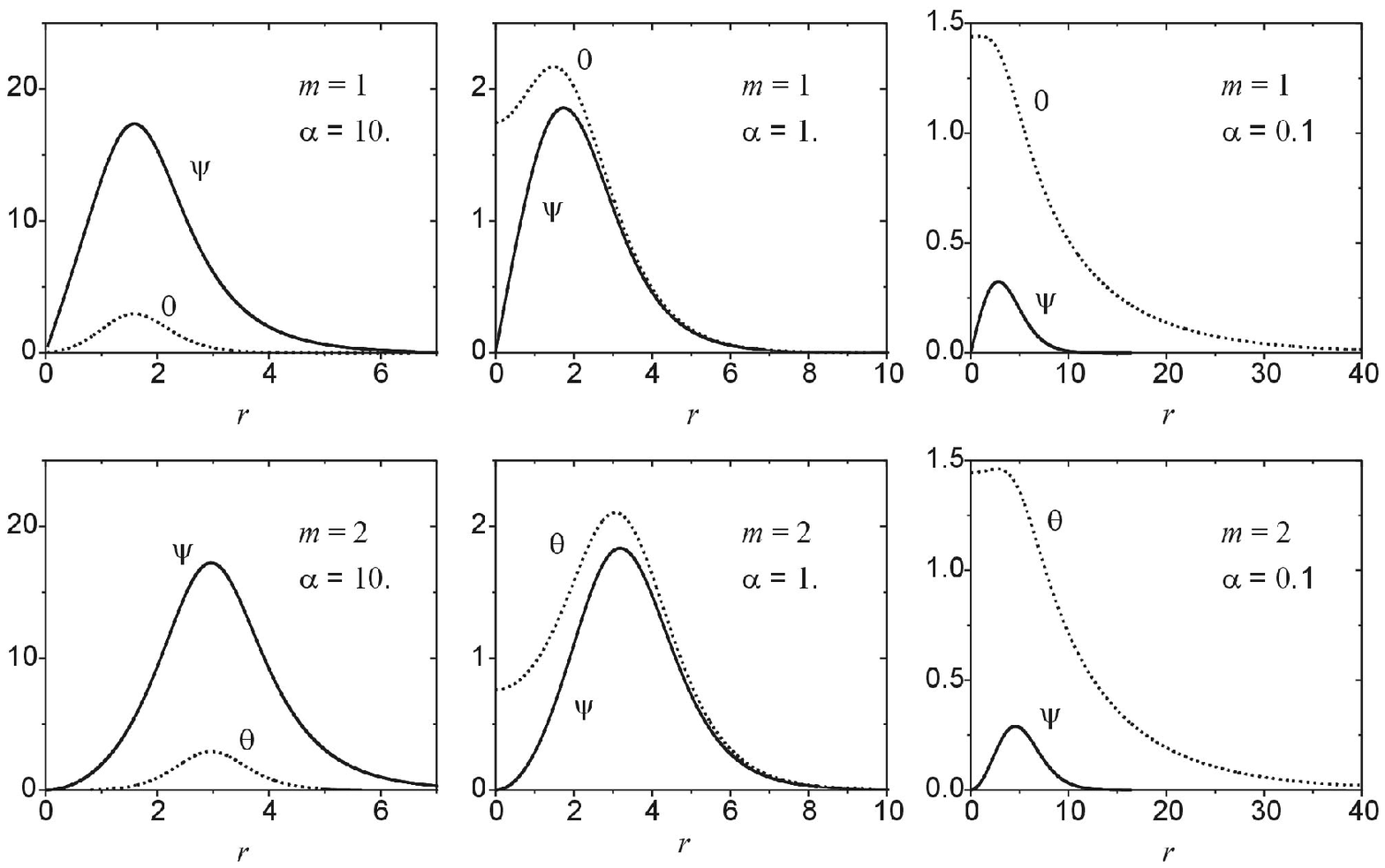}
\caption{Examples of the stationary vortex solutions for $m=1$ and
$m=2$ at different values of the nonlocality parameter $\alpha$.
Shown are the fields $\psi(r)$ (solid) and $\theta(r)$ (dotted).}
\label{fig:Cuts}
\end{figure}

\begin{figure}
\includegraphics[width=3.4in]{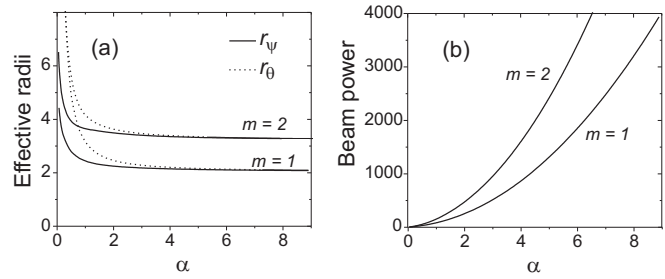}
\caption{ (a) Effective radii of the field intensity distribution
$r_\psi$ (solid)  and the temperature field $r_\theta$ (dotted)
vs. the nonlocality parameter $\alpha$, for $m=1,2$.
 (b) Power $P$ vs. $\alpha$.} \label{fig:EffectiveAll}
\end{figure}

We solve the nonlinear integral equation (\ref{eq:E_rintegralEq})
using the stabilized relaxation procedure similar to that employed
in Ref.~\cite{Petviashvili86}. Figure \ref{fig:Cuts} shows several
examples of the solutions of the system (\ref{eq:psi_r}) found
numerically for different values of nonlocality parameter
$\alpha$. To characterize these solutions, we define the effective
radii $r_\psi$ and $r_\theta$ of the intensity distribution
$\left| \psi \right|^2$ and the temperature perturbation
distribution $\theta$, respectively, as follows,
$$r^2_\psi=\frac1N\int r^2|\psi(r)|^2d^2 \textbf{r}, \;\;\;
 r^2_\theta=\frac{\int r^2\theta(r)d^2 \textbf{r}}{\int\theta(r)d^2
\textbf{r}}.
$$
Figure  \ref{fig:EffectiveAll}(a) shows the radii $r_\psi$ and
$r_{\theta}$ as functions of the nonlocality parameter $\alpha$.
Both $r_\psi$ and $r_\theta$ decrease monotonically when the
nonlocality parameter grows. In the local limit ($\alpha\gg 1$),
$r_\psi$ and $r_\theta$ saturate at the same finite value, which
increases with the topological charge.
Figure~\ref{fig:EffectiveAll}(b) shows the beam power $P =
\int\left|E \right|^2 d^2 \textbf{r}$ as a function of the
nonlocality parameter $\alpha$.

The important information on stability of the vortex solitons can
be obtained from the analysis of small perturbations of the
stationary states. The basic idea of such a linear stability
analysis is to represent a perturbation as the superposition of
the modes with different azimuthal symmetry. Since the
perturbation is assumed to be small, stability of each linear mode
can be studied independently. Presenting the nonstationary
solution in the vicinity of the stationary mode as follows,
\[
 \Psi(\textbf{r} ,z)=
\left\{\psi(r)+\varepsilon_+(r)e^{i\omega z+i L\varphi}
 +\varepsilon_-^*(r)e^{-i\omega^* z-i L\varphi}\right\}e^{i\lambda
 z}, \]
 \[
 \Theta(\textbf{r},z)=\theta(r)+\vartheta_+(r)e^{i\omega z+i L\varphi}
 +\vartheta_-^*(r)e^{-i\omega^* z-i L\varphi},
\]
where $|\varepsilon_\pm|\ll \psi$, $|\vartheta_\pm|\ll\theta$,
$\psi$, $\theta$ are assumed to be real without loss of
generality, we linearize Eqs.~(\ref{eq:NLS}) and obtain the system
of linear equations of the form:
 \begin{equation}
 \label{eq:EigenProblem}
   \pm\left\{-\lambda+
\Delta_r^{(m\pm L)}+\theta(r)+\hat
g_L\right\}\varepsilon_\pm\pm\hat g_L
   \varepsilon_\mp=\omega\varepsilon_\pm,
 \end{equation}
where $$\hat g_L\varepsilon_\pm=\psi(r)\int_0^{\infty}\xi
\psi(\xi)G_{L}(r,\xi; \alpha)\, \varepsilon_{\pm}(\xi)d\xi.
$$

\begin{figure*}
\includegraphics[width=\textwidth]{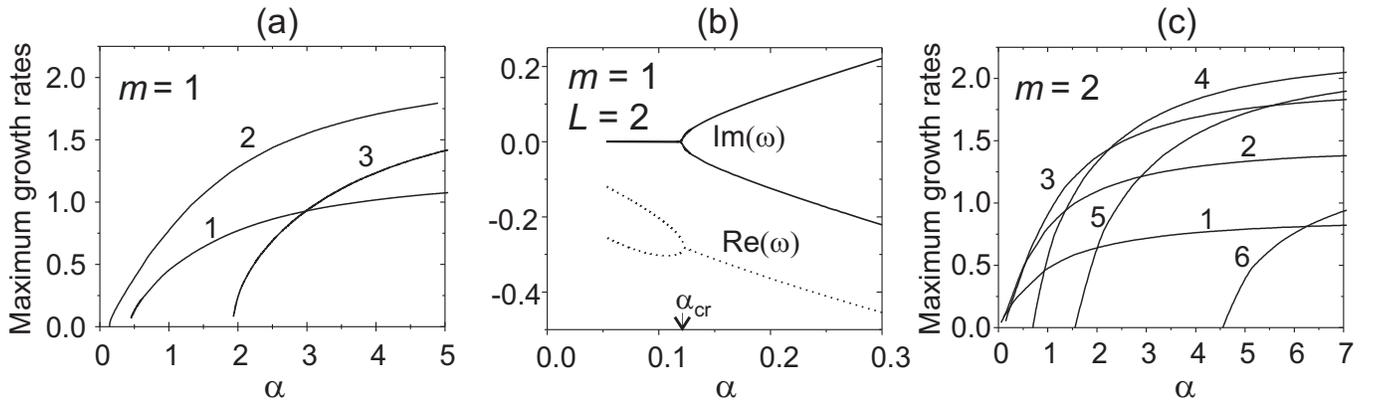}
 \caption{Maximum
growth rate of linear perturbation modes vs. the nonlocality
parameter $\alpha$ for the vortices with (a) $m=1$ and (c) $m=2$.
Numbers near the curves stand for the azimuthal mode numbers $L$.
(b) Real and imaginary parts of the eigenvalue $\omega$ of the
most dangerous azimuthal mode with $L=2$ and $m=1$.}
\label{fig:GrRates}
\end{figure*}

The Hankel spectral transform was applied to reduce the
integro-differential eigenvalue problem (\ref{eq:EigenProblem}) to
linear algebraic one. The maximum growth rate $|{\rm Im}\,
\omega|$ of linear perturbation modes is shown in
Fig.~\ref{fig:GrRates}(a) for a single-charge  vortex ($m=1$). The
symmetry-breaking modes can become unstable only for $L=1,2,3$.
The growth rates saturate in the local regime $\alpha\gg 1$. The
largest growth rate as well as the widest instability region has
the azimuthal mode with the number $L=2$. The real and imaginary
parts of the eigenvalues $\omega$ for this most dangerous mode are
shown in Fig.~\ref{fig:GrRates}(b). Importantly, there exists a
bifurcation point $\alpha_\textrm{cr}\approx 0.12$ below which the
growth rate $\textrm{Im}\, (\omega)$ vanishes.  Thus, the
symmetry-breaking azimuthal instability is eliminated in a
highly-nonlocal regime: all growth rates vanish provided
$\alpha<\alpha_\textrm{cr}$.

We also perform the linear stability analysis for multi-charge
vortices with the topological charges $m=2$ and $m=3$.
Figure~\ref{fig:GrRates}(c) shows the growth rate of the linear
perturbation modes for the vortex with $m=2$. Importantly, the
growth rate of the $L=2$ mode always remains nonzero, and the same
result holds for the vortices with $m=3$. Therefore, the linear
stability analysis predicts the existence of stable single-charge
vortex in a highly nonlocal regime, while the multi-charge
vortices are shown to be unstable with respect to decay into the
fundamental solitons, even in the limit $\alpha\to 0$.

\begin{figure}
\includegraphics[width=3.4in]{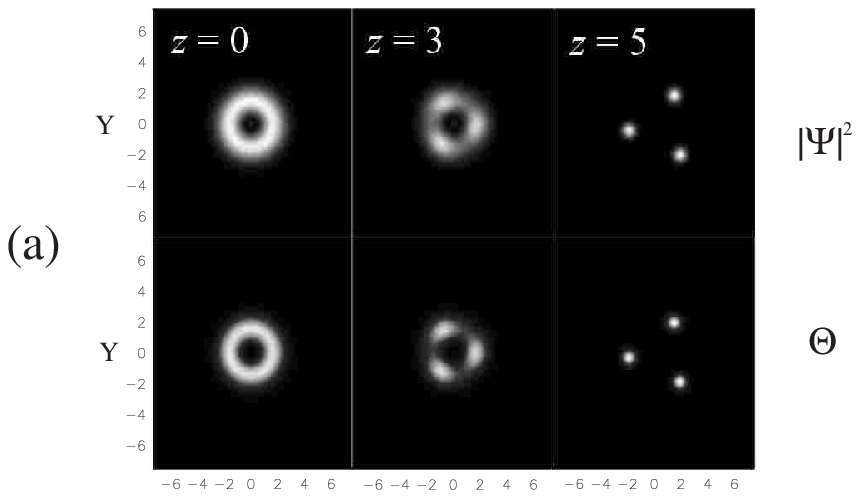}\\

\includegraphics[width=3.4in]{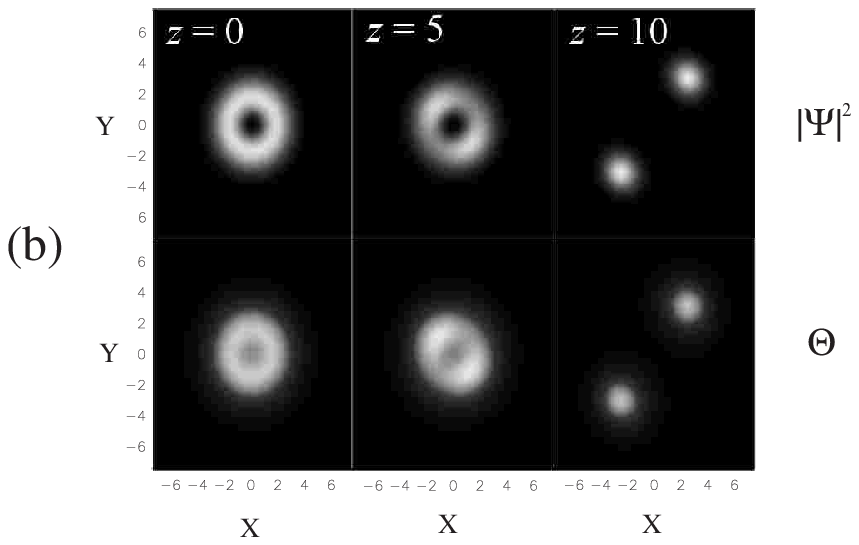}
 \caption{ Evolution of
the beam intensity $|\Psi|^2$ (upper row) and temperature field
$\theta$ (lower row) of a perturbed single-charge vortex soliton
for (a) $\alpha=5$ and (b) $\alpha=1$.} \label{fig:Decay}
\end{figure}

\begin{figure}
\includegraphics[width=3.4in]{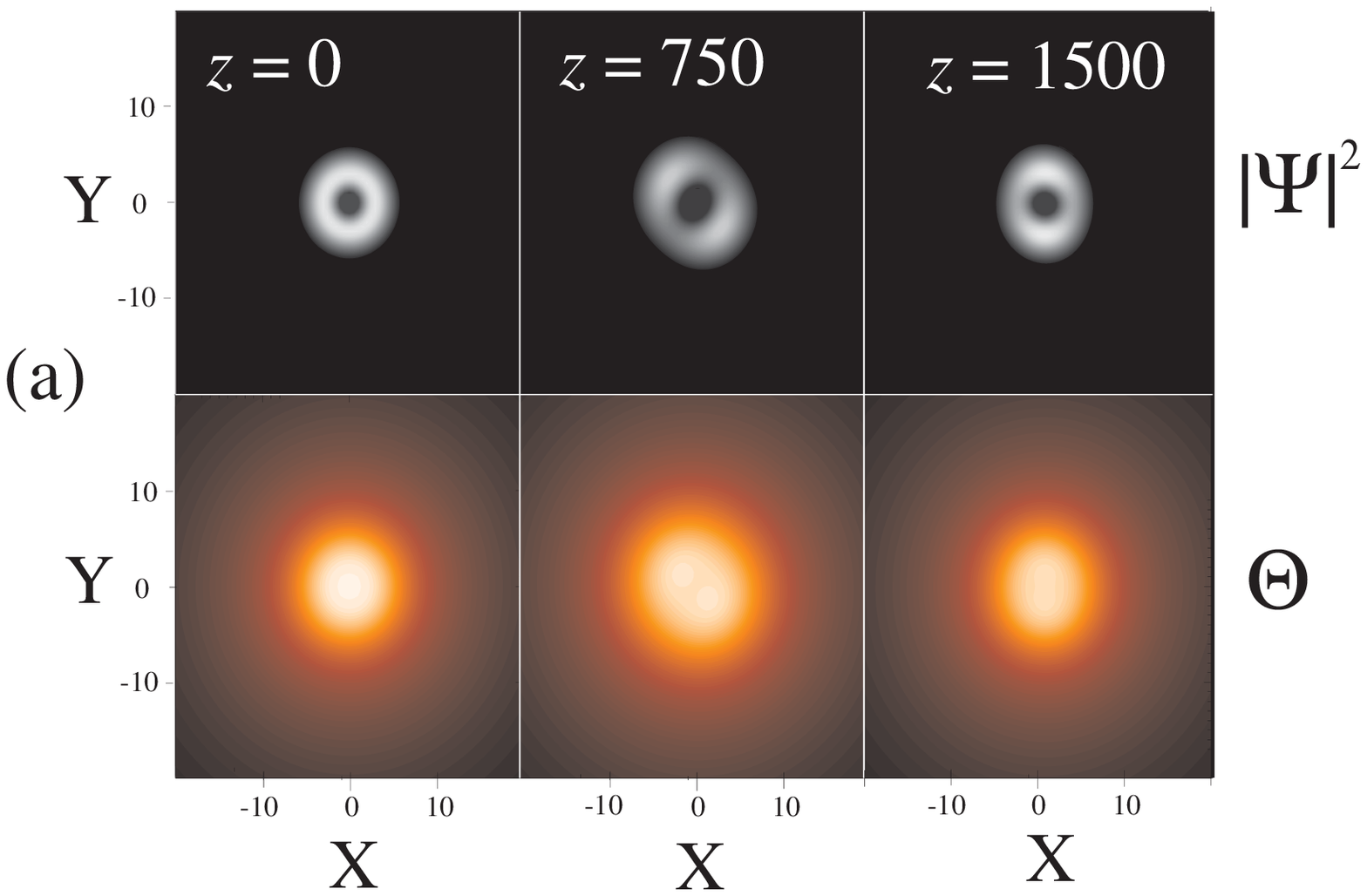}\\

\includegraphics[width=3.4in]{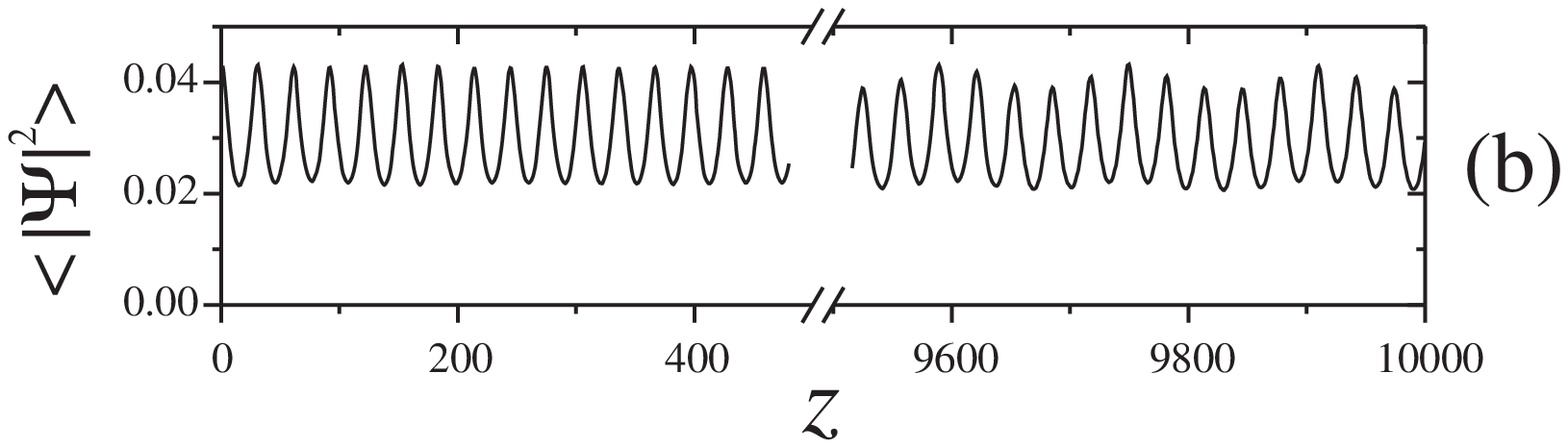}
\caption{ (a) Stable propagation of a single-charged vortex
generated by a singular Gaussian beam with $m=1$ ($h=0.14,$
$w=4.2$, $\alpha=0.07$) shown for the beam intensity $|\Psi|^2$
and temperature field $\theta$. (b) Oscillatory dynamics of the
amplitude of the intensity field $\langle|\Psi|^2\rangle$  vs. $z$
during the beam propagation. }
 \label{fig:GaussStart}
\end{figure}

Our stability analysis has been verified with direct simulations
of the propagation dynamics of perturbed vortex solitons by
employing the split-step Fourier method to solve
Eqs.~(\ref{eq:NLS}) numerically. The results of our linear
stability analysis agree well with the numerical simulations. In
particular, the symmetry-breaking instabilities have been observed
in the region predicted by the linear stability analysis, and some
examples of the vortex decay instability are presented in
Fig.~\ref{fig:Decay} for the vortices with $m=1$. If a
perturbation is applied to a single-charge vortex in the strongly
nonlocal regime (such that all azimuthal instabilities are
completely suppressed by the nonlocality), the vortex beam evolves
in a quasi-periodic fashion: the effective radii and amplitudes
oscillate with $z$. Thus, our numerical simulations indicate that
single-charge vortex solitons become stable if the nonlocality
parameter $\alpha$ is below some critical value which is very
close to the value $\alpha_\textrm{cr} \approx 0.12$ predicted by
the linear stability analysis. In experiments, the input beam may
differ essentially from the exact stationary solution. Therefore,
we perform additional numerical simulations for singular Gaussian
input beams of the form $\Psi(r,0)= h r\exp(-r^2/w^2+i\varphi)$
and, as follows from Fig.~\ref{fig:GaussStart}, observe that such
beams are indeed stabilized below the critical value of
nonlocality, and the beam effective intensity, defined as
$\langle|\Psi|^2\rangle=N^{-1}\int|\Psi|^4d^2 \textbf{r}$,
underdoes large-amplitude oscillations.

The physical mechanism for suppressing the symmetry-breaking
azimuthal instability of the vortex beam in a nonlocal nonlinear
medium can be understood as being associated with effective
diffusion processes introduced by a nonlocal response. Indeed, if
a small azimuthal perturbation of the radially-symmetric vortex
deforms its shape in some region, the corresponding temperature
distribution along the vortex ring becomes nonuniform. As a
result, the intrinsic thermodiffusion processes would smooth out
this inhomogeneity and suppress its further growth, leading to the
complete vortex stabilization in a highly-nonlocal regime.

In conclusion, we have studied the basic properties and stability
of spatially localized vortex beams in a self-focusing nonlinear
medium with a nonlocal response. We have found that single-charge
optical vortices can be stabilized with respect to any
symmetry-breaking azimuthal instability in the regime of strong
nonlocal response, whereas multi-charged vortices remain unstable
and decay into the fundamental solitons for any degree of
nonlocality. Our numerical simulations confirm that the stable
propagation of single-charge vortex is indeed possible in a
nonlocal nonlinear medium. We expect that these results will
stimulate the first experimental observation of stable vortices in
self-focusing nonlinear media.

\newpage



\newpage
\begin{thebibliography}{99}

\bibitem{book} L.M. Pismen, {\em Vortices in Nonlinear Fields} (Clarendon Press,
Oxford, 1999), and references therein.

\bibitem{review} M.S. Soskin and M.V. Vasnetsov, in {\em Progress in Optics}, edited
by E. Wolf (North-Holland, Amsterdam, 2001), Vol. 42, p. 219.

\bibitem{KivsharAgrawal} See, e.g., Yu.S. Kivshar and G. Agrawal,
{\em Optical Solitons: From Fibers to Photonic Crystals} (Academic
Press, San Diego, 2003), Sec. 6.5 and references therein.

\bibitem{theory} M. Quiroga-Teixeiro and H. Michinel, J. Opt. Soc. Am. B
{\bf 14}, 2004 (1997); A. Desyatnikov, A. Maimistov, and B.A.
Malomed, Phys. Rev. E {\bf 61}, 3107 (2000); H. Michinel, J.
Campo-T\'aboas, M.L. Quiroga-Teixeiro, J.R. Salqueiro, and R.
Grac\'ia-Fern\'andez, J. Opt. B {\bf 3}, 314 (2001); V. Skarka,
N.B. Aleksi\'c, and V.I. Berezhiani, Phys. Lett. A {\bf 291}, 124
(2001); I. Towers, A.V. Buryak, R. A. Sammut, B. A. Malomed, L.-C.
Crasovan, and D. Mihalache, Phys. Lett. A {\bf 288}, 292 (2001);
Phys. Rev. E {\bf 63}, 055601 (2001); B.A. Malomed, L.-C.
Crasovan, and D. Mihalache, Physica D {\bf 161}, 187 (2002); D.
Mihalache, D. Mazilu, L.-C. Crasovan, I. Towers, A.V. Buryak, B.
A. Malomed, L. Torner, J. P. Torres, and F. Lederer, Phys. Rev.
Lett. {\bf 88}, 073902 (2002); T.A. Davydova, A.I. Yakimenko, and
Yu.A. Zaliznyak, Phys. Rev. E {\bf 67}, 026402 (2003); T.A.
Davydova and A.I. Yakimenko, J. Opt. A {\bf 6}, 197 (2004); D.
Mihalache, D. Mazilu, B.A. Malomed, and F. Lederer, Phys. Rev. E
{\bf 69}, 066614 (2004).

\bibitem{remark} Stabilization of vortices
created by {\em partially incoherent light} was recently observed
in a self-focusing photorefractive crystal, see C.-C. Jeng, M.-F.
Shih, K.  Motzek, and Yu.S. Kivshar, Phys. Rev. Lett. {\bf 92},
043904 (2004).

\bibitem{LitvakSJPlPhys75} A.G. Litvak, V.A. Mironov, G.M.
Fraiman, and A.D. Yunakovskii, Sov. J. Plasma Phys. {\bf 1}, 60
(1975).

\bibitem{ContiPRL03} C. Conti, M. Peccianti, and G. Assanto, Phys.
Rev. Lett. {\bf 91}, 073901 (2003).

\bibitem{BEC} See, e.g., L. Santos, G.V. Shlyapnikov, and M. Lewenstein,
Phys. Rev. Lett. {\bf 90}, 250403 (2003).

\bibitem{TuritsynTMF85} S.K. Turitsyn, Theor. Math. Phys. {\bf 64}, 226 (1985).

\bibitem{KrolikovskiJOptB04} W. Kr{\'o}likowski, O. Bang, N.I. Nikolov, D. Neshev,
J. Wyler, J.J. Rasmussen, and D. Edmundson, J. Opt. B {\bf 6}, 288
(2004).

\bibitem{ContiPRL04} C. Conti, M. Peccianti, and G. Assanto, Phys.
Rev. Lett. {\bf 92}, 113902 (2004).

\bibitem{Petviashvili86} V.I. Petviashvili and V.V. Yan'kov, Rev.
Plasma Phys. Vol. 14, Ed. B.B. Kadomtsev,  (Consultants Bureau,
New York, 1989), pp 1-62.
\end{thebibliography}
\end{document}